\documentclass[12pt,oneside]{article}

\date{20th August 2018}

\usepackage[ruled,linesnumbered]{algorithm2e}
\usepackage{color}
\usepackage[a4paper, total={7in, 9in}]{geometry}
\usepackage{graphicx}		
\usepackage{amsmath}
\usepackage{caption}
\usepackage{color}
\usepackage{amssymb}

\begin{document}
\title{Matrix Completion with Weighted Constraint for Haplotype Estimation}
\author{S. Majidian, M. Mohades, and M.H. Kahaei}
\maketitle

\section{abstract}
A new optimization design is proposed for matrix completion by weighting the measurements and deriving the corresponding error bound. Accordingly, the Haplotype reconstruction using nuclear norm minimization with Weighted Constraint (HapWeC) is devised for haplotype estimation. Computer simulations show  the outperformance of the HapWeC compared to some recent algorithms in terms of the normalized reconstruction error and reconstruction rate. 

\section{Introduction}
Matrix completion has already been applied to collaborative filtering, system identification, global positioning, and  remote sensing problems. A model defined for matrix completion is \cite{can}
\begin{equation}
Y_{ij}= M_{ij}+ Z_{ij} \quad \forall (i,j)\in \Omega
\end{equation}
where $Y_{ij}, M_{ij}$, and $Z_{ij}$ are the entries of  $\boldsymbol Y$, $\boldsymbol M$, and $\boldsymbol Z$, respectively showing the measurement, desired low-rank, and noise matrices, all with $N\times l$ dimensions. Also,  $\Omega$ represents the measurement set and, without loss of generality, we assume that $N<l$. To estimate $\boldsymbol M$, the following minimization has been proposed \cite{can}.
\begin{equation}
\min_{\boldsymbol X}{\| \boldsymbol X \|_*}  \quad \textrm {s.t. } \| P_\Omega(\boldsymbol X - \boldsymbol Y) \|_F \leq \delta
\end{equation}
in which $[P_\Omega(\boldsymbol A)]_{ij}=A_{ij}$ for $(i,j)\in \Omega$ and zero, otherwise. Also, $\|\cdot\|_F$ and  $\|\cdot\|_*$ denote the Frobenius and the nuclear norms, respectively.

Here, we consider the haplotype reconstruction problem, $a.k.a$ haplotype  assembly problem \cite{si,maj} in which the quality of each measurement is defined by $Q_{ij}$. Then, the error probability of the $(i,j)^{th}$ measurement; which is exploited to estimate the haplotypes more accurately, is given by $P_{ij}=10^{-Q_{ij}/10}$ \cite{ill}.

Here, we first propose a new weighted optimization scheme in which each measurement is utilized based on its $Q_{ij}$ and the corresponding error bound is derived. Accordingly, the weights  are optimized using $Q_{ij}$'s. At last, an algorithm is developed to estimate haplotypes.
\section{Proposed optimization}
In order to cope with diverse quality of data, we introduce the following optimization problem called the Nuclear norm minimization with a Weighted Constraint (NuWeC):
\begin{align}
\min_{\boldsymbol X}{\| \boldsymbol X \|_*}  \quad \textrm {s.t. } \|\boldsymbol W  \odot P_\Omega(\boldsymbol X - \boldsymbol Y) \|_F \leq \delta,
\end{align}
where $\odot$ is the Hadamard product and $\boldsymbol W$ is the weight matrix which will be introduced in the next sections.  The geometric interpretation of proposed optimization in (3) is illustrated in Fig. 1 in which the ellipsoid is the feasible set intersecting the smallest nuclear norm ball at $\hat{\boldsymbol M}$  showing the optimal point. The error bound of the NuWeC is derived in Theorem 1.
\vspace*{-0.2cm}
\begin{figure}[h]
\centering{\includegraphics[width=50mm]{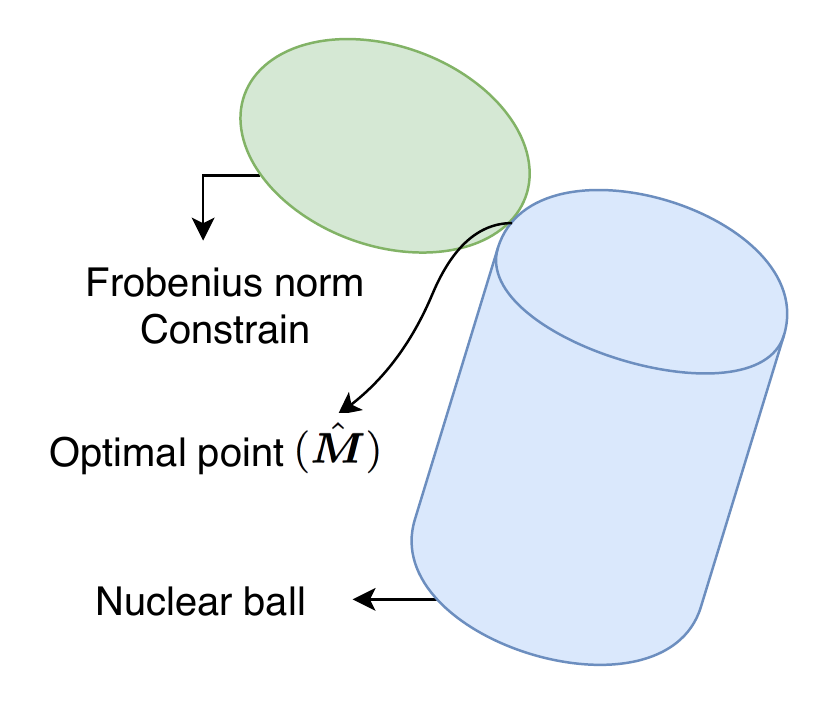}}
\caption{Geometric interpretation of the proposed nuclear norm minimization with weighted constraint.}
\end{figure}

\textbf{Theorem 1.} Consider $\hat{\boldsymbol M}$  as the optimal point of the optimization problem (3). Then, we obtain
\begin{align}
\| \hat{\boldsymbol M} - \boldsymbol M \|_F  \leq  2\delta \sqrt{ \frac{p+2}{p}\frac{N}{(1-\alpha)^2} +1 } \left \{ \sum_{ij} \frac{1}{W_{ij}^2} \right \}^{1/2} 	
\end{align}
in which $p=\frac{|\Omega|}{Nl}$ is the sampling rate and $0<\alpha<1$ is a numerical constant.\\
\textbf{Proof:} By denoting $\boldsymbol H\triangleq \hat{\boldsymbol M}-\boldsymbol M$, we intend to bound  $\| \boldsymbol H \|_F^2= \| \boldsymbol H_\Omega\|_F^2+\| \boldsymbol H_{\Omega^c} \|_F^2$,  where $\boldsymbol H_{\Omega}= P_\Omega (\boldsymbol H)$, $\boldsymbol H_{\Omega^c}= P_{\Omega^c} (\boldsymbol H)$, and $P_{\Omega^c}$ is the complement operator of $P_{\Omega}$. One can easily see that the following inequality holds,
\begin{align}
\| \boldsymbol H_\Omega\|_F \leq \|P_\Omega(\hat{\boldsymbol M} - \boldsymbol Y)\|_F +\| P_\Omega(\boldsymbol M - \boldsymbol Y)\|_F.
\end{align}
To bound the term $\|P_\Omega(\hat{\boldsymbol M} - \boldsymbol Y)\|_F$, we first note that for a give matrix $\boldsymbol{A} $, using the Holder inequality and   $\|\boldsymbol{a}\|_2 \leq \|\boldsymbol{a} \|_1$, we can derive
\begin{align}
\| \boldsymbol{A} \|_F  &\leq \left \{ \sum_{ij} \frac{1}{W_{ij}^2} \right \}^{1/2}  \left \{ \sum_{ij} W_{ij}^2 A_{ij}^4  \right \}^{1/2}  \nonumber\\ &\leq \left \{ \sum_{ij} \frac{1}{W_{ij}^2} \right \}^{1/2}   \sum_{ij} W_{ij} A_{ij}^2.
\end{align}
Then, for the feasibile point $\hat{\boldsymbol M}$ in (3) which satisfies the constraint $\|\boldsymbol W  \odot P_\Omega(\hat{\boldsymbol M} - \boldsymbol Y) \|_F \leq \delta$ and defining $\boldsymbol{A}= P_\Omega(\hat{\boldsymbol M} - \boldsymbol Y)$ , we obtain from (6)
\begin{align}
\| P_\Omega(\hat{\boldsymbol M} - \boldsymbol Y)\|_F    \leq
\left \{ \sum_{(i,j)\in \Omega} \frac{1}{W_{ij}^2} \right \}^{1/2}\delta.
\end{align}
Now, we show the feasibility of $\boldsymbol M$ in (3) to conclude that (7) also holds for $\boldsymbol M$ similar to $\hat{\boldsymbol M}$. To do so, we can write
\begin{align}
\|\boldsymbol W  \odot P_\Omega(\boldsymbol M - \boldsymbol Y) \|_F^2 & = \sum_{(i,j)\in \Omega} W_{ij}^2(M_{ij}-Y_{ij})^2 \nonumber \\& =\sum_{(i,j)\in \Omega} W_{ij}^2 Z_{ij}^2 \leq  \|\boldsymbol W\|_{\infty}^2\|\boldsymbol Z\|_F^2 <\delta^2,
\end{align}
where $\|\boldsymbol W\|_{\infty}=\max{\boldsymbol W_{ij}}$.  This result shows that $\boldsymbol M $  is feasible for $\delta>\|\boldsymbol W\|_{\infty}  \|\boldsymbol Z\|_F $ and thus the last term of (5) is  bounded.
Using these results in (5) leads to
 \begin{align}
\| \boldsymbol H_\Omega\|_F \leq
2\delta \left \{ \sum_{(i,j)\in \Omega} \frac{1}{W_{ij}^2} \right \}^{1/2}. 	
\end{align}
On the other hand, based on \cite{can}, with a high probability, $\boldsymbol H_{\Omega^c}$ obeys
\begin{align}
\| \boldsymbol H_{\Omega^c}\|_F^2 \leq  (1+\frac{2}{p})\frac{N}{(1-\alpha)^2} \| \boldsymbol H_\Omega\|_F^2,
\end{align}
in which $0<\alpha<1$ can be taken equal to $\frac{1}{2}$. From (9) and (10), the bound given by (4) in Theorem 1 is proved.

\section{Optimization of weights}
We now consider the bound derived in Theorem 1 as an objective function to optimize $\boldsymbol{W}$ as follows:
\begin{align}
\min_{\boldsymbol W}{   2\delta \sqrt{ \frac{p+2}{p}\frac{N}{(1-\alpha)^2} +1 } \left \{ \sum_{ij} \frac{1}{W_{ij}^2} \right \}^{1/2}}\textrm{ s.t. }\| P_\Omega(\boldsymbol W)\|_{\infty}=1.
\end{align}
Furthermore, in order to exploit the error probabilities, we suggest the following relationship:
\begin{align}
W_{ij}= a\log_{2}{(\frac{1}{P_{ij}})}+b \quad (i,j) \in \Omega,
\end{align}
in which an entry with a lower error probability will be more effective on the penalty term of (3), $i.e.$, $\sum W_{ij}^2(M_{ij}-Y_{ij})^2$.
Making use of the logarithmic function enables us to incorporate all the measurements while restricting the large variation of  error values.
Then, by substituting (12) in (11), we get the following optimization problem:
\begin{align}
\min_{a,b}{\sum{\frac{1}{(a\log_{2}{(\frac{1}{P_{ij}})+b)^2}} }}  \quad \textrm{s.t. }   \max\{a\log_{2}{(\frac{1}{P_{ij}})+b}\}=1.
\end{align}
Using $b=1+a\log_{2}{P_{\textrm{min}}} $ in (13), the corresponding unconstrained non-convex optimization  problem may be solved by a grid search.
\section{Proposed algorithm for haplotype reconstruction}
For the haplotype reconstruction problem, $\boldsymbol Y \in \{0,\pm1\}^{N\times l}$, $\boldsymbol M\in \{\pm1\}^{N\times l}$, and $\boldsymbol Z\in \{0,\pm2\}^{N\times l}$ described by (1) are the read, haplotype, and the noise matrices, respectively \cite{si}. For diploids, $\boldsymbol M$ consists of two different rows $\boldsymbol h_1$ and $\boldsymbol h_2$  and thus its rank is 2. The goal of haplotype reconstruction is to estimate two rows of $\boldsymbol M$ using the read matrix. By exploiting the NuWec optimization  problem given by (3),  we develop the  "Haplotype reconstruction using nuclear norm minimization with Weighted Constraint (HapWeC)" algorithm as below.
\begin{algorithm}
\SetAlFnt{\small}
\SetAlgoLined
\SetNoFillComment
\SetKwInOut{Input}{Input}\SetKwInOut{Output}{Output}
\Input{$N$ reads and quality scores $Q_{ij}$}
\Output{Haplotypes}
Construct the read matrix $\boldsymbol M$ ($N\times l$).\\
Calculate the error probability using
$P_{ij}=10^{-Q_{ij}/10}$.\\
Find the weights based on (13).\\
Find $\hat{\boldsymbol M}$ using convex optimization problem (3).\\
Compute the SVD of $\hat{\boldsymbol M}=\sum_{i=1}^{r}{\sigma_ix_iy_i^T}$.\\
Truncate the SVD by setting all singular values to zero except the two largest ones: $T_2(\hat{\boldsymbol M})=\sum_{i=1}^{2}{\sigma_ix_iy_i^T}$.\\
Obtain haplotypes $\hat{\boldsymbol h_1}$ and $\hat{\boldsymbol h_2}$  by extracting the independent rows of $T_2(\hat{\boldsymbol M})$.\\
Round the haplotypes $\hat{\boldsymbol h}_1$ and $\hat{\boldsymbol h}_2$ to $\pm 1$.\\
\caption{Haplotype reconstruction using nuclear norm minimization with Weighted Constraint (HapWeC)}
\end{algorithm}

It can be shown that by truncating the Singular Values Decomposition (SVD) of $\hat{\boldsymbol M}$,  the error bound is changed by a factor of  $k=1+\sqrt{\textrm{rank} (\hat{\boldsymbol{M}})+1 }$, $i.e.$, $ \| T_2(\hat{\boldsymbol M}) - \boldsymbol M \|_F^2\leq k\| \hat{\boldsymbol M} - \boldsymbol M \|_F^2$. 
 \section{Simulation results}
First, we evaluate the NuWeC using a synthetic dataset.  To do so, a rank-two random matrix $\boldsymbol M \in \{\pm 1\}^{40\times 40}$ is generated   whose 10\% of entries are contaminated with noise. We consider both nuclear minimization problem and the NuWeC defined by (2) and (3), respectively. The Normalized Reconstruction Error (NRE) is defined as
\begin{align}
NRE=\frac{1}{n} \sum_{i=1}^{n} \frac{\| \hat{\boldsymbol M}^{(i)}- \boldsymbol M \|_F }{\|\boldsymbol M \|_F},
\end{align}
where $\hat{\boldsymbol M}^{(i)}$ shows the estimated desired matrix in the $i^{th}$ experiment and $n=20$ is the number of independent Monte Carlo experiments.
The NREs are shown as a function of the sampling percentage in Fig. 2. As seen, the NREs  of NuWeC decreases about 2dB which is effectively due to incorporation of the quality scores.
\begin{figure}[!ht]
\centering{\includegraphics[width=100mm]{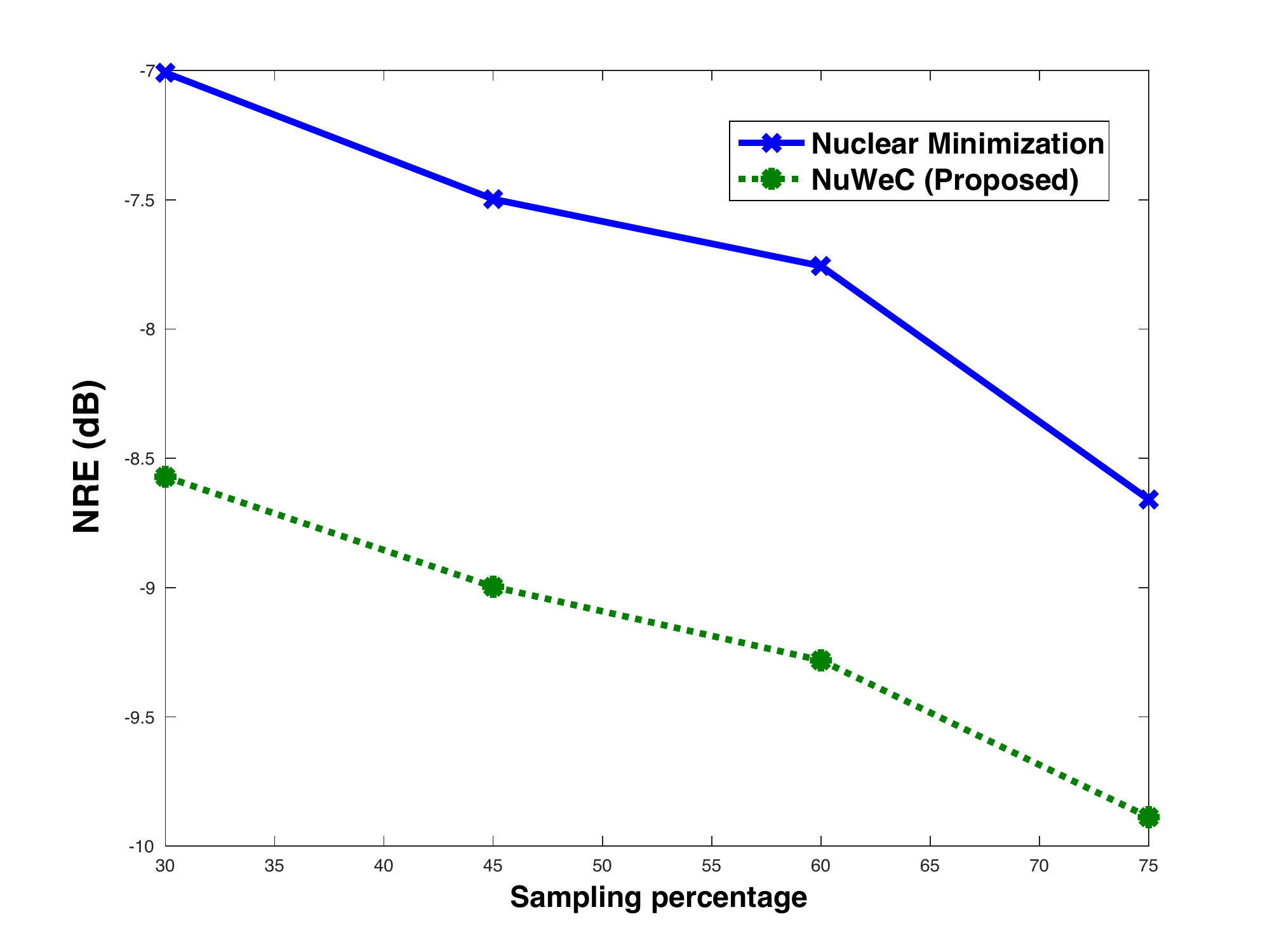}}
\caption{NREs of NuWeC and nuclear minimization problems  vs. the sampling percentage for the synthetic data.}
\end{figure}

In the second scenario, we consider the read database of \cite{ger}.  The number of reads and the haplotype length  are selected as $N=86$ and $l=100$, respectively. Also, the sampling percentage is $p=7\%$ and the coverage per column is 6. The results in Fig. 3 show the superiority of the NuWeC compared to the nuclear minimization by reducing the NREs.
\begin{figure}[!ht]
\centering{\includegraphics[width=100mm]{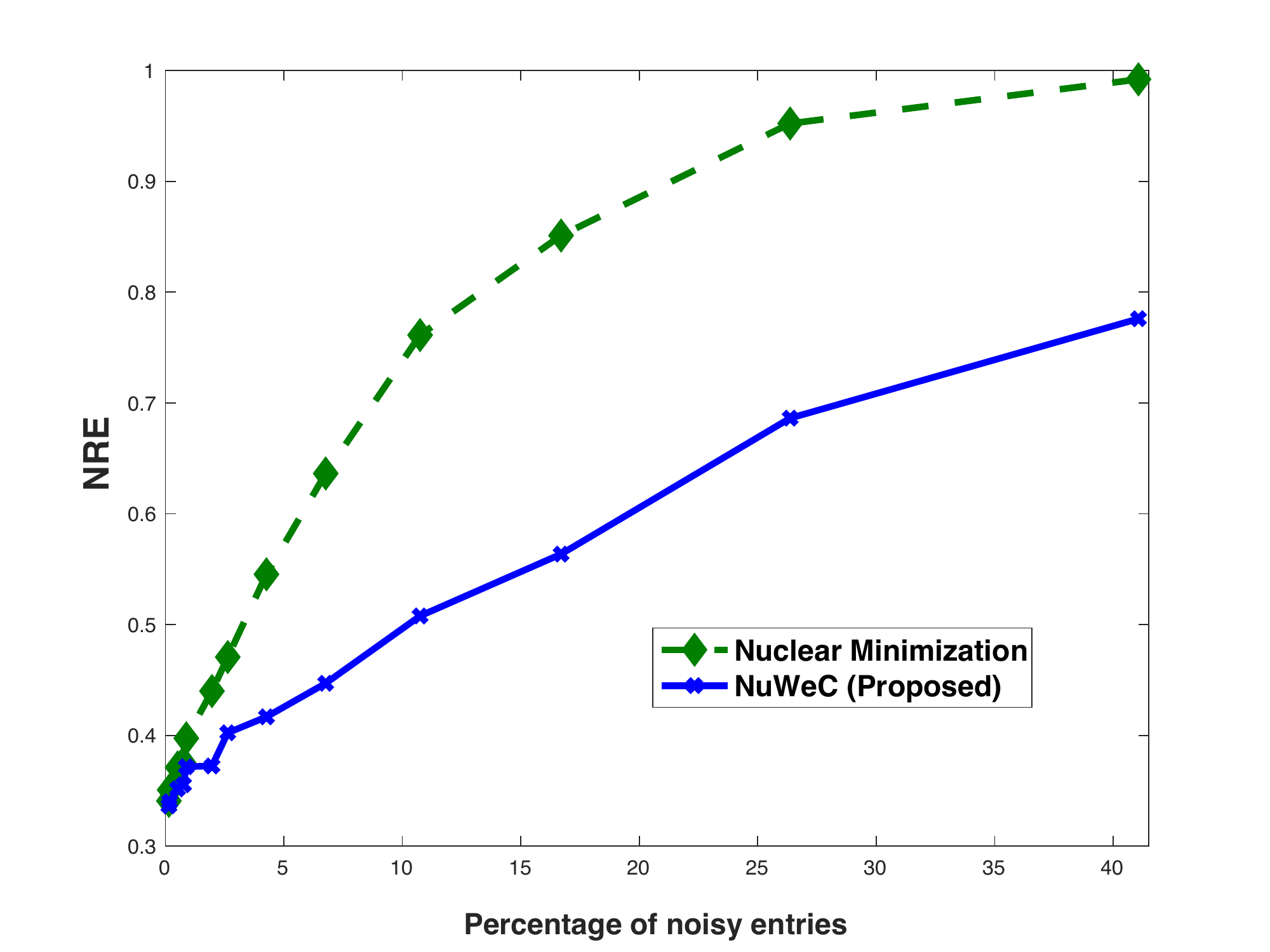}}
\caption{NREs vs. the percentage of noisy entries for the NuWeC and nuclear minimization problems based on the dataset addressed in \cite{ger}.}
\end{figure}
Now, we compare the proposed HapWeC  with the nuclear minimization, NuWeC, and alternating minimization algorithm \cite{cai} for haplotype reconstruction. To inspect the estimated and actual haplotypes, the  reconstruction rate (rr) is defined as \cite{ger}
\begin{align}
rr=1 -\frac{1}{4nl} \sum_{i=1}^{n}  \| \hat{\boldsymbol h}_1^{(i)}-\boldsymbol h_1\|_1 +\| \hat{\boldsymbol h}_2^{(i)}-\boldsymbol h_2\|_1,
\end{align}
where $\hat{\boldsymbol h}_2^{(i)}$ and $\hat{\boldsymbol h}_1^{(i)}$
are the estimated haplotypes of the $i^{th}$ experiment.
Decrease of the reconstruction rates shown in Fig. 4  reveal the outperformance of the developed HapWeC.
\begin{figure}[!ht]
\centering{\includegraphics[width=100mm]{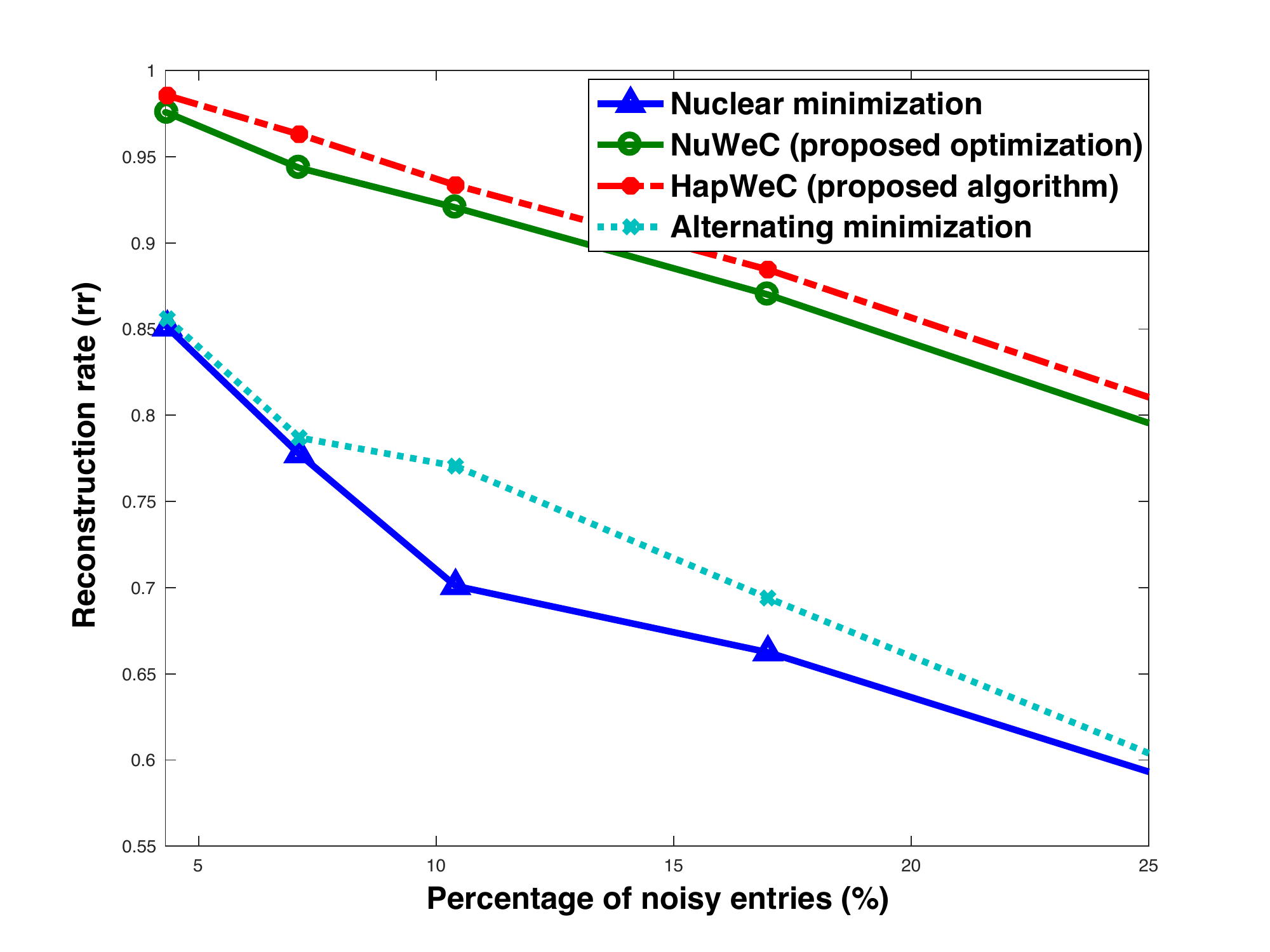}}
\caption{Comparison of reconstruction rate  vs. the percentage of noisy entries for different algorithms  based on the dataset addressed in \cite{ger}.}
\end{figure}
\section{Conclusion}
The NuWec, a new weighted optimization algorithm was developed for matrix completion by exploiting the quality of measurements and the corresponding error bound was derived. Computer simulations showed about 2dB reduction in the resulting estimation error compared to that of the nuclear norm minimization technique.  The NuWeC was then used to design the new HapWeC algorithm for haplotype estimation. This algorithm increased the reconstruction rate about 10\% in camparison to  some recent methods.
\vskip5pt 
\noindent Mohammad Hossein Kahaei, (School of Electerical Engineering, Iran University of Science \& Technology, Tehran, Iran.)
\vskip3pt
\noindent E-mail: kahaei@iust.ac.ir


\begin{thebibliography}{}
\bibitem{can}
Candes, E. J.,  Plan, Y. (2010). Matrix completion with noise. Proceedings of the IEEE, 98(6), 925-936.
\bibitem{ger}
Geraci, F. (2010). A comparison of several algorithms for the single individual SNP haplotyping reconstruction problem. Bioinformatics, 26(18), 2217-2225.
\bibitem{si}
Si, H., Vikalo, H.,  Vishwanath, S. (2017). Information-theoretic analysis of haplotype assembly. IEEE Transactions on Information Theory, 63(6), 3468-3479.
\bibitem{ill}
Illumina Inc. Quality scores for next-generation sequencing. Technical report, 2011.
\bibitem{maj}
Majidian, S.,  Kahaei, M. H. (2018). NGS Based Haplotype Assembly Using Matrix Completion. arXiv preprint arXiv:1801.09864.
\bibitem{cai}
Cai, C., Sanghavi, S.,  Vikalo, H. (2016). Structured low-rank matrix factorization for haplotype assembly. IEEE Journal of Selected Topics in Signal Processing, 10(4), 647-657.
\end{thebibliography}
\end{document}